\documentclass[reprint,aps,prl,superscriptaddress,nobalancelastpage]{revtex4-2}

\usepackage{graphicx}
\usepackage[dvipsnames]{xcolor}
\usepackage{amsmath,amssymb}
\usepackage{ulem}
\definecolor{darkblue}{rgb}{0,0,0.6}
\definecolor{darkred}{rgb}{0.6,0,0}
\usepackage[colorlinks=true,urlcolor=darkblue, citecolor=darkblue, linkcolor=darkred, hyperfootnotes=false]{hyperref}

\begin{document}

\title{Universal wrinkling dynamics of a sheet on viscous liquid}

\author{Ayrton Draux}
\affiliation{Laboratoire Interfaces $\&$ Fluides Complexes, Universit\'e de Mons, 20 Place du Parc, B-7000 Mons, Belgium.}
\author{Marco Rizzo}
\affiliation{Laboratoire Interfaces $\&$ Fluides Complexes, Universit\'e de Mons, 20 Place du Parc, B-7000 Mons, Belgium.}
\author{Dominic Vella}
\affiliation{Mathematical Institute, University of Oxford, Woodstock Rd, Oxford OX2 6GG, UK.}
\author{Vincent D\'emery}
\affiliation{Laboratoire de Physique, Univ Lyon, École Normale Supérieure de Lyon, CNRS, Lyon 69342, France}
\affiliation{Gulliver UMR CNRS 7083, ESPCI Paris, PSL Research University, 10 rue Vauquelin, 75005 Paris, France.}
\author{Fabian Brau}
\email[]{fabian.brau@ulb.be}
\affiliation{Universit\'e libre de Bruxelles (ULB), Nonlinear Physical Chemistry Unit, CP231, 1050 Bruxelles, Belgium}
\author{Pascal Damman}
\email[]{pascal.damman@umons.ac.be}
\affiliation{Laboratoire Interfaces $\&$ Fluides Complexes, Universit\'e de Mons, 20 Place du Parc, B-7000 Mons, Belgium.}


\date{\today}

\begin{abstract}
We investigate the wrinkling dynamics of a thin elastic sheet that is indented or compressed while floating on a viscous liquid. We show that the deformation speed controls the dynamics, leading to a wrinkle wavelength significantly smaller than that selected under quasistatic compression. Once active compression ceases, the wrinkles coarsen until their wavelength relaxes toward the equilibrium value. We develop a theoretical model coupling Stokes flow in the liquid to elastic bending of the sheet, which quantitatively predicts both the initial wavelength selection and its subsequent coarsening. We demonstrate that the same mechanism governs two-dimensional and axisymmetric geometries, thereby extending classical static wavelength selection laws to dynamic situations. Although developed from controlled laboratory experiments, the model captures a generic viscous–elastic coupling and applies broadly to thin elastic films interacting with viscous environments, including the formation of surface wrinkles in \emph{p\={a}hoehoe} lava flows.
\end{abstract}

%\keywords{}

\maketitle

\textit{Introduction} -- Thin sheets easily deform when subjected to an applied force. They typically buckle when compressed~\cite{landau1986theory} and may crumple so that the stress focused in localized regions is separated by isometric domains~\cite{BenAmar1997,Witten2007,Tallinen2009,Cambou2011,Timounay2020}. In contrast, regular wrinkles appear when a tensile or shear force is applied~\cite{Cerda2002,Wong2006}. Thin films can be attached to a soft substrate and also develop wrinkles under compression. 
For a liquid substrate of density $\rho$, the wrinkles are  sinusoidal for small quasistatic compression with a wavelength given by $\lambda_{S} = 2\pi (B / \rho g)^{1/4}$, where $B=Eh^3/12(1-\nu^2)$ is the bending stiffness with $E$, $h$, $\nu$ the Young modulus, thickness and Poisson ratio of the film ($g$ being the gravitational acceleration)~\cite{Milner1989,cerda2003geometry,brau2013wrinkle}. At larger compression, the deformation is localized in a single deep fold~\cite{pocivavsek2008,brau2015wrinkling}. Besides measuring specific properties of very thin films~\cite{Stafford2004,Chung2011}, these extremely regular patterns have found applications in, e.g., optics to create tunable phase gratings~\cite{Harrison2004}, improve light extraction from organic light-emitting diodes~\cite{Koo2010}, increase the efficiencies of photovoltaic cells~\cite{Kim2012} and in stretchable electronics~\cite{Khang2006,Rogers2010}.
Another way to generate regular wrinkles is the indentation of a sheet floating on a fluid. Here, the compression is indirect: for sufficiently large indentations, the radial motion of the sheet towards the indenter produces an azimuthal compression. According to the idea of asymptotic isometry, this compressive stress relaxes through the growth of radial wrinkles, characterized again by their wavelength, which depends on the distance to the indenter and converges to $\lambda_S$ far from it~\cite{holmes2010,vella2015indentation,vella2018regimes,paulsen2016,ripp2020geometry}.

Few works have studied the compression/indentation at a finite speed of a sheet resting on a liquid. Indentation at high speed can be realized by the impact of a solid sphere on a stretchable~\cite{Vandenberghe2016} or inextensible~\cite{box2019dynamics,Kiely2020} sheet floating on water. This dynamics at large Reynolds numbers allows the viscous effects to be neglected to compute the coarsening of the wavelength with time. The effects of viscosity $\mu$ on the wavelength of the wrinkles and relaxation dynamics are not yet clearly established. Indeed, it has been investigated using uniaxially compressed films~\cite{chatterjee2015wrinkling,Wang2023} and indented sheets~\cite{Wang2023} but different scalings have been proposed for the wavelength $\lambda_0$ that emerges initially ($\lambda_0 \sim  V^{-1/2}$~\cite{chatterjee2015wrinkling} versus $\lambda_0 \sim  V^{-1/6}$~\cite{Wang2023}) as well as for the subsequent coarsening of the wrinkles with time ($\lambda\sim t^{1/4}$ \cite{Im2005,Im2006} versus $\lambda\sim t^{1/6}$ \cite{vandeparre2010hierarchical,Kodio2017}). Note also that when the compression speed and thickness $H$ of the liquid layer are small enough, ridges may appear instead of wrinkles~\cite{guan2022compression,guan2023compression}.

\begin{figure*}[!t]
\includegraphics[width=\textwidth]{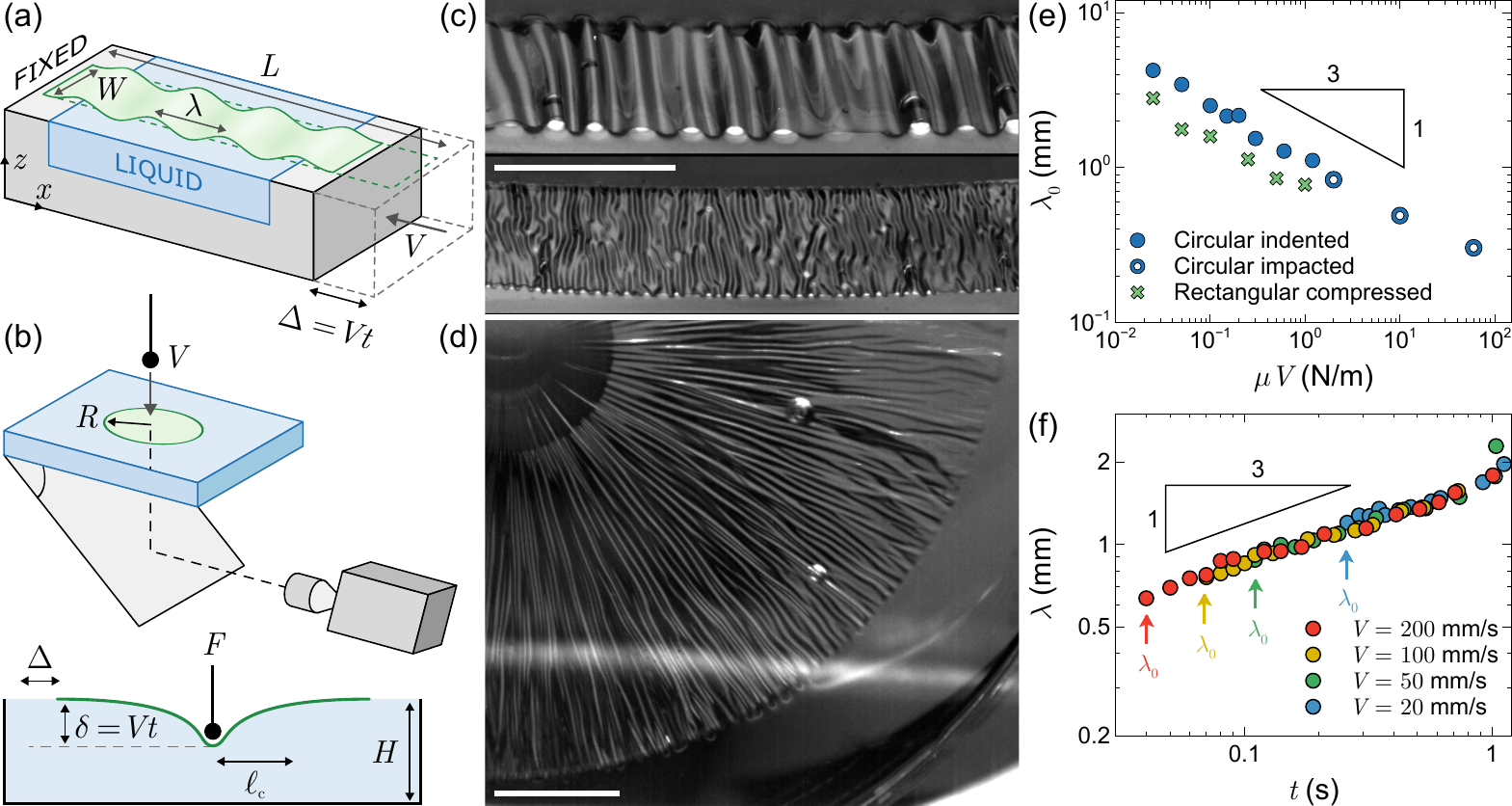}
\caption{Schematic of the experimental setup used to (a) compress or (b) indent/impact  a thin PET/PDMS sheet floating on a viscous liquid. (c) Wrinkles on a rectangular PDMS sheet ($h = 50$ $\mu$m) floating on silicon oil ($\mu = 5$ Pa s) compressed at $V = 5$ mm/s (top) and $V= 200$ mm/s (bottom). (d) Wrinkles on a PET circular sheet ($R=35$ mm and $h = 3$ $\mu$m) floating on a fluid ($\mu = 5$ Pa s) at $t=3$ ms after being impacted by a marble at $V=2$ m/s. Scale bars in (c)-(d): 1 cm. (e) Dependence of the initial wavelength $\lambda_0$ on $\mu V$ for a circular PET sheet ($h=3$ $\mu$m, $R=63$ mm) resting on silicon oils ($1 \leq \mu \leq 30$ Pa s) and impact or indented at speeds raging from 5 mm/s to 2 m/s. The rectangular PDMS sheet ($h=50$ $\mu$m, $L=39$ mm) is resting on silicon oil ($\mu = 5$ Pa s) and confined at speeds from 5 mm/s to 200 mm/s. (f) Coarsening of the wrinkled pattern for a rectangular PDMS thin sheet ($h=50$ $\mu$m) kept at a given compression ratio on a viscous fluid ($\mu=30$ Pa s) after being compressed at various speeds as indicated ($\lambda_S \simeq 7$ mm).  
}
\label{fig01}
\end{figure*}

In this Letter, we study the formation of wrinkles on a thin sheet floating on a thick layer of liquid of viscosity $\mu$ and compressed or indented at a speed $V$. We show that the initial wavelength of the wrinkles is determined by $\mu V$ and can be drastically shorter than its equilibrium value before it slowly coarsens towards it. We develop a theoretical model that couples the Stokes equations and the beam equation to determine the initial wavelength and its coarsening dynamics. The compression/indentation of a sheet on a viscous fluid gives an exquisite control of the wrinkle pattern that gives access to a wavelength much smaller than the equilibrium value, $\lambda_{S}$. 

\textit{Experiments} -- The setups used to study the uniaxial confinement of rectangular sheets and the indentation of circular sheets floating on a fluid are shown schematically in Fig.~\ref{fig01}(a,b). Rectangular sheets of length $L$ and width $W$, as well as circular sheets of radius $R$, were made from spin-coated polydimethylsiloxane (PDMS, $h=50$ or $75$ $\mu$m, $E \simeq 1$ MPa) or polyethylene terephthalate (PET, $3 \le h \le 15$ $\mu$m, $E \simeq 5$ GPa) and shaped with a laser cutter. Indentation experiments were performed with a constant indentation speed $V$ varying between $5$ and 40 mm/s using sheets of radius $35 \leq R \leq 63$ mm resting on a bath of liquid (viscosity $1 \leq \mu \leq 30$ Pa s, density $\rho =970$ kg/m$^3$) of depth $H=10$ or 25 mm. Uniaxial compression experiments were achieved at constant speeds $V$ ranging from $0.1$ to $200$ mm/s thanks to a motorized stage (Oriental Motor) and using sheets with $39 \leq L \leq 69$ mm and $10 \leq W \leq 20$ mm. Additional impact experiments with steel balls were also used to show the universal character of the observations.

\textit{Results} --
Figure~\ref{fig01}(c,d) shows typical wrinkling patterns observed for the uniaxial compression of a rectangular sheet and the indentation of a circular sheet, respectively. Figure~\ref{fig01}(c) shows the drastic influence of the compression speed on the wrinkle wavelength that decreases when $V$ increases. These observations are rationalized in Fig.~\ref{fig01}(e) showing that the evolution of the initial wavelength, $\lambda_0$, with $V$ does not depend on the setup used: uniaxial compression, indentation, or impact. In all cases, its value decreases with the compression/indentation speed as $\lambda_0\sim (\mu V)^{-1/3}$. Figure~\ref{fig01}(f) shows that when the compression ratio, $\Delta/L$, is kept constant after the emergence of the wrinkles during a rapid uniaxial confinement, their wavelength slowly increases over time following a power law, $\lambda(t) \sim t^{1/3}$. This relaxation dynamics does not depend on the initial compression speed as seen in Fig.~\ref{fig01}(f) where the data collapse well when the compression rate is varied by one order of magnitude. We note however that this power law is only valid when $\lambda \ll \lambda_S$ since coarsening must stop when $\lambda = \lambda_S$.

\begin{figure}[!t]
\centering
\includegraphics[width=\columnwidth]{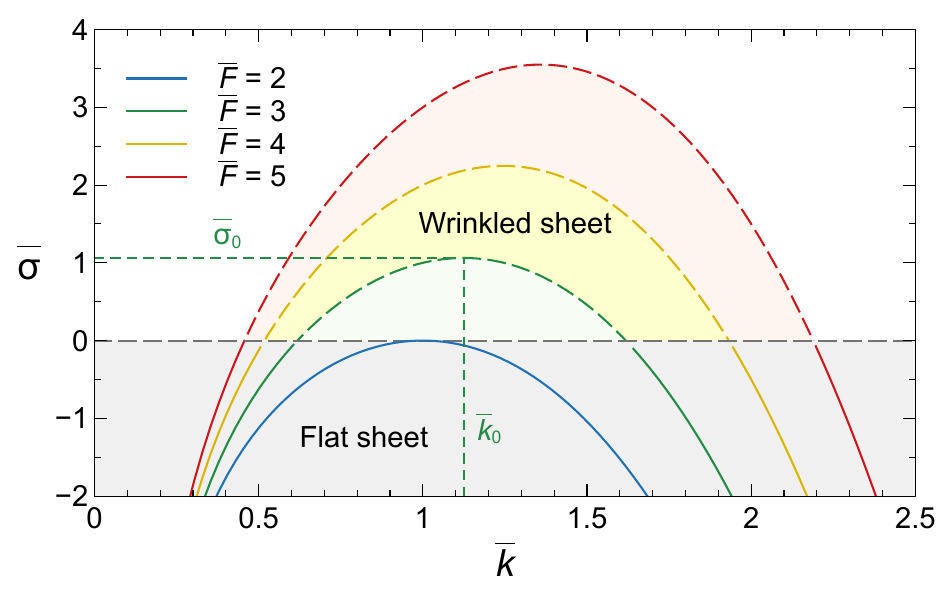}
\caption{Evolution of the wrinkle growth rate $\bar{\sigma}={2\mu \sigma}/{B k_S^3}$ as a function of the wavenumber $\bar k =k/k_S$ for various $\bar{\mathcal{F}}={\mathcal{F}}/{B k_S^2}$. With this notation, Eq.~(\ref{disp-rel-main}) becomes $\bar{\sigma} = -\bar{k}^3 + \bar{\mathcal{F}} \bar{k} - {1}/{\bar{k}}$ (see \cite{SM}). The most unstable wavenumber $\bar{k}_{0}$ and its corresponding growth rate $\bar{\sigma}_{0}$ are shown for $\bar{\mathcal{F}}=3$.}
\label{fig02}
\end{figure}

\begin{figure}
    \centering  \includegraphics[width=\columnwidth]{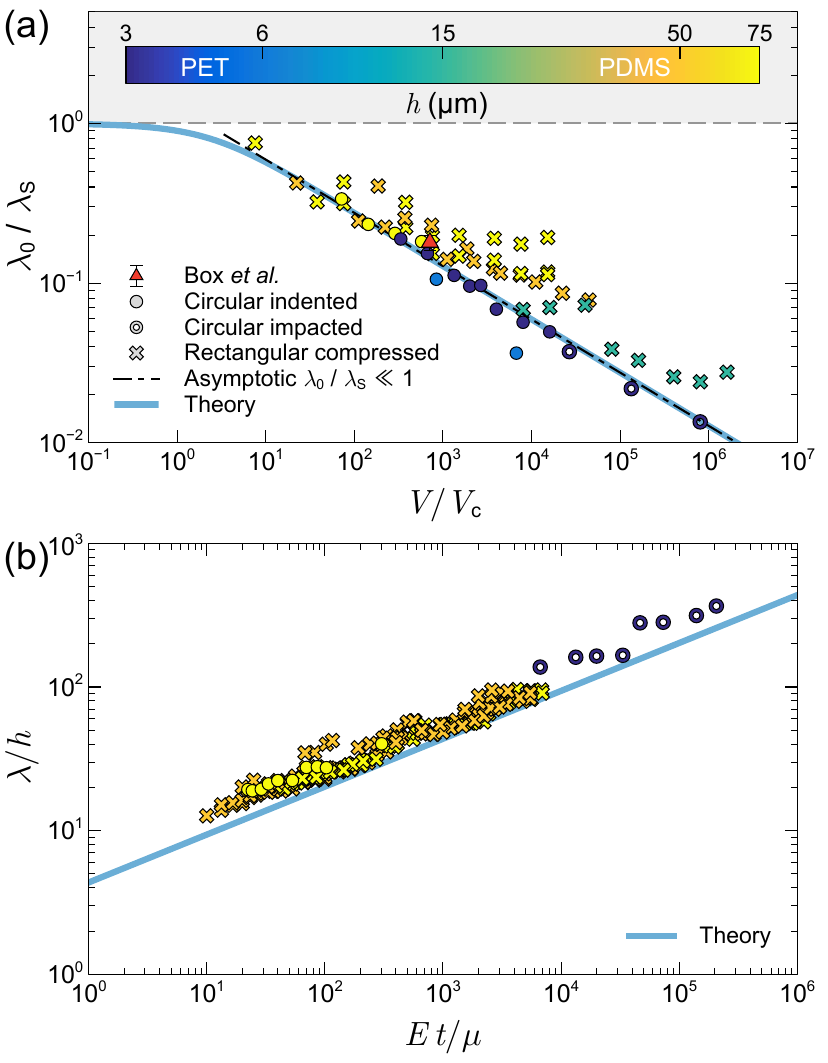}
    \caption{(a) Evolution of the rescaled initial wavelength, $\lambda_0 /\lambda_S$, for different experimental setups as a function of the rescaled compression/indentation speed, $V/V_c$, where $V_c$ is defined in Eq.~(\ref{lambda0}) for rectangular sheets and by Eq.~(\ref{vc_circular}) for circular sheets. The solid curve is obtained by solving numerically Eq.~(\ref{lambda0}) whereas the dashed-dotted curve corresponds to the asymptotic power-law (\ref{lambda0-asymp}). The data from Box \textit{et al.} are also shown~\cite{Box2017}. (b) Evolution of the rescaled wavelength, $\lambda/h$, as a function of the rescaled time, $E t/\mu$, during the coarsening regime. The solid curve corresponds to Eq.~(\ref{lambda-coarsening}) with $\nu = 0.5$.}
    \label{fig03}
\end{figure}

\textit{Theoretical model} -- The observed dynamics of the wavelength results from the coupling between the mechanical stress related to the growth of wrinkles at imposed compression ratio and the viscous drag flows induced by the sheet motion. To derive a model, we consider a rectangular sheet of length $L \gg \lambda$ floating on a liquid and compressed at a constant speed $V$. The 2D Stokes equations in the plane ($x$, $z\le w(x,t)$) where $w(x,t)$ is the sheet profile are given by:
\begin{equation}
\label{eqs-u-main}
\mu \nabla^2 \vec{u} =\vec{\nabla} \bar{p}, \quad \vec{\nabla}\cdot \vec{u}=0,
\end{equation}
%\begin{subequations}
%\label{eqs-u-main}
%\begin{align}
%&\frac{\partial^2 u_x}{\partial x^2} + \frac{\partial^2 u_x}{\partial z^2} = \frac{1}{\mu} \frac{\partial \bar{p}}{\partial x}, \quad \frac{\partial^2 u_z}{\partial x^2} + \frac{\partial^2 u_z}{\partial z^2} = \frac{1}{\mu} \frac{\partial \bar{p}}{\partial z}, \\
%&\frac{\partial u_x}{\partial x} + \frac{\partial u_z}{\partial z} = 0,
%\end{align}
%\end{subequations}
where $\vec{u}=(u_x,u_z)$ and $\bar{p} = p + \rho g z$ with $p$ the pressure. These equations are solved with the boundary conditions (BCs) $u_x = u_z =0,\ \bar{p}=0$ for $z \to -\infty$.
The coupling of viscous flows to the sheet deformation is obtained via the pressure given by the beam equation, $\bar{p} = B \partial_x^4 w + \mathcal{F} \partial^2_x w + \rho g w$,  at $z = w(x,t)$, where $\mathcal{F}$ is the compressive force per unit width, which we assume constant for simplicity. Note that we consider an infinitely deep liquid container since velocity fields vanish when $|z|$ is much larger than $\lambda$ and $H\gg\lambda$ in the experiments. Far enough from the extremities, the
sheet profile should be given by $w(x, t) = A(t) \cos(kx)$, where the amplitude $A$ is small compared to its wavelength at the onset of wrinkling, $wk \to 0$. The Stokes equations imply that $\nabla^2 \bar{p}=0$ which, together with the two BCs above for $\bar{p}$, yields
\begin{equation}
\label{p-sol-main}
\bar{p} = \alpha(k)\, A(t) \cos(kx) e^{kz}, \quad \alpha = Bk^4-\mathcal{F} k^2+\rho g.
\end{equation}
Using Eq.~(\ref{p-sol-main}) in Eq.~(\ref{eqs-u-main}) together with the BCs above for the fluid velocity $\vec{u}$, leads to~\cite{SM}
\begin{subequations}
\begin{align}
\label{u-sol-main}
u_x &= \frac{\alpha A }{2 \mu}(w -z) \sin(k x) \, e^{kz}, \\
u_z &= -\frac{\alpha A}{2\mu k} [1+k(w-z)]\cos(k x) \, e^{kz}.
\end{align}
\end{subequations}
Imposing that the sheet stays attached to the liquid, $u_z = \partial_t w$ at $z = w(x,t)$ and with $wk\to 0$, we obtain the dynamical equation for the wrinkle amplitude
\begin{equation}
\label{eq-amp-wrink-main}
\frac{d A(t)}{dt} = -\frac{\alpha(k)\, A(t)}{2\mu k},
\end{equation}
whose solution reads as
\begin{equation}
\label{disp-rel-main}
A(t) = A_0 \, e^{\sigma\, t},\quad \sigma =  -(Bk^4-\mathcal{F} k^2+\rho g)/(2\mu k).
\end{equation}
As shown in Fig.~\ref{fig02}, the wavenumber $k_0$ maximizing the wrinkle growth rate is given by $\partial_k \sigma|_{k_0}=0$ for a given force $\mathcal{F}$ which yields the relation,
\begin{equation}
\label{rel-F-k-main}
\mathcal{F} = 3Bk_0^2 - \rho g k_0^{-2}.
\end{equation}
 Substituting this last relation in Eq.~(\ref{disp-rel-main}) yields the growth rate of the wrinkle amplitude as a function of the selected wavenumber, $k_0$.
\begin{equation}
\label{sigma0-main}
\sigma_0 = (Bk_0^3- \rho g k_0^{-1})/\mu.
\end{equation}
The dispersion relation (\ref{disp-rel-main}) shows that wrinkled states ($\sigma >0$) exist only for $\mathcal{F} > \mathcal{F}_c= 2 (B \rho g)^{1/2}$ which corresponds to the static wrinkling threshold~\cite{pocivavsek2008,brau2015wrinkling}. The largest imposed displacement $\Delta_c$ the sheet can sustain before  wrinkling is given by $\mathcal{F}_c = \bar{E} h \Delta_c/L$ leading to $\Delta_c  = 2 (B \rho g)^{1/2} L/(\bar{E}h)$, where $\bar{E} = E/(1-\nu^2)$. The total compression ratio writes $\Delta = \Delta_c + V t$ so that $t=0$ at the onset of wrinkling. Assuming inextensibility in the wrinkled state~\cite{pocivavsek2008,brau2011multiple,brau2015wrinkling}, we have
\begin{equation}
\label{inex-main}
\frac{\Delta}{L} = \frac{A^2 k_0^2}{4} \quad \Rightarrow \quad \frac{\Delta_c}{L} + \frac{V t}{L} = \frac{A_0^2 k_0^2}{4} (1+2\sigma_0 t),
\end{equation} 
where we used the expression (\ref{disp-rel-main}) of $A$ expanded near $t=0$. Inextensibility thus implies that the rate of growth, for $t$ close to 0, is given by $\sigma_0 = V/(2\Delta_c)$. Equating this last expression to Eq.~\eqref{sigma0-main}, we get the expression of $k_0$ in terms of the control parameters of the system $Bk_0^3- \rho g k_0^{-1} = \mu V/(2\Delta_c)$. 
This last relation can be written in terms of $\bar{\lambda}_0 = \lambda_0/\lambda_S$ with $\lambda_0 = 2\pi / k_0$ as follows
\begin{equation}
\label{lambda0}
\bar{\lambda}_0^{-3}-\bar{\lambda}_0 = \frac{3}{2\pi }\frac{V}{V_c}, \quad V_c = \frac{3}{\pi}\frac{Bk_S^3 \Delta_c}{\mu}=\frac{\rho g L h^2}{\mu \lambda_S}
\end{equation}
where we used the expression of $\Delta_c$ given above and $k_S = 2\pi/\lambda_S$. Note that, when $V \ll V_c$, Eq.~\eqref{lambda0} becomes $\bar{\lambda}_0^{-3}-\bar{\lambda}_0 \simeq 0$, showing that the initial wavelength tends to the static one, i.e.~$\bar{\lambda}_0 \to 1$. When $V \gg V_c$, $\bar{\lambda}_0$ becomes small so that the first term of the left-hand side of Eq.~(\ref{lambda0}) dominates and $\bar{\lambda}_0$ is then given by a power-law
\begin{equation}
\label{lambda0-asymp}
\frac{\lambda_0}{\lambda_S} = \left[\frac{2\pi }{3}\right]^{1/3} \left[\frac{V}{V_c}\right]^{-1/3}.
\end{equation}
A similar relation can be derived for indented circular sheets where $V$ is the indentation speed and
\begin{equation}
\label{vc_circular}
    V_c = \frac{3}{2\pi} \frac{B k_S^3 \delta_c}{\mu} = \frac{\sqrt{3}}{2^{4/3}\pi^{5/3}}\frac{\rho g h R}{\mu}\left[\frac{B}{\rho g R^4}\right]^{1/12},
\end{equation}
with $\delta_c$ the indentation depth beyond which wrinkles appear~\cite{SM}. The numerical solution of Eq.~(\ref{lambda0}) together with the asymptotic expression (\ref{lambda0-asymp}) are shown in Fig.~\ref{fig03}(a) for the different geometries. They are in very good agreement, without any fitting parameter, with the observed initial wavelength, including previously published data obtained in the early stages of impacts for sheet on low viscosity fluids~\cite{box2019dynamics}. Moreover, the result appears to be completely independent of the method used to compress the sheet: uniaxial compression or indentation/impact. We emphasize that $\lambda_0$ is determined by an interplay between bending, gravitational and viscous forces, which precludes its derivation from a simple dimensional analysis. Further, this result is distinct from the local $\lambda$-law used to describe static wrinkle patterns~\cite{paulsen2016}.

As seen in Fig.~\ref{fig01}(f), the wrinkle pattern coarsens when a rectangular sheet is kept confined at a fixed displacement $\Delta$, the wavelength $\lambda$ growing as $t^{1/3}$ as long as $\lambda \ll \lambda_S$. During this coarsening, we still have a wrinkled profile so that Eq.~(\ref{eq-amp-wrink-main}) and inextensibility (\ref{inex-main}), i.e. ${\Delta}/{L} = {A^2 k(t)^2}/{4}$, still apply. Using this last expression to substitute $A$ in favor of $k$ in Eq.~(\ref{eq-amp-wrink-main}) yields the evolution equation for $k$,
\begin{equation}
\label{k-ODE-main}
\frac{d k(t)}{dt} = \frac{Bk(t)^4-\mathcal{F} k(t)^2+\rho g}{2\mu},
\end{equation}
where we used the expression (\ref{p-sol-main}) of $\alpha$. Assuming that at each time the growth rate is maximum, the relation (\ref{rel-F-k-main}) between the force and the most unstable wavenumber still applies, with $k_0$ replaced by $k(t)$,  implying that $\mathcal{F}$ varies in time. In this case, Eq.~(\ref{k-ODE-main}) becomes
\begin{equation}
\label{k-ODE-final-main}
\frac{d k(t)}{dt} = \left[-B k(t)^4+\rho g \right]/\mu.
\end{equation}
The stationary solution of Eq.~(\ref{k-ODE-final-main}) is $k = k_S$ -- the quasi-static wavenumber. When $k \gg k_S$, Eq.~(\ref{k-ODE-final-main}) admits a power-law solution, $k = (3Bt/\mu)^{-1/3}$, which can be written as
\begin{equation}
\label{lambda-coarsening}
\frac{\lambda(t)}{h} = \frac{2\pi}{[4(1-\nu^2)]^{1/3}} \left(\frac{E t}{\mu}\right)^{1/3},
\end{equation}
where we used the explicit expression of $B$. This relation is in good quantitative agreement with the observed evolution of $\lambda$ with time (Fig.~\ref{fig03}(b)). We note that Eq.~(\ref{lambda-coarsening}) is valid at early times and breaks down when $k\simeq k_S$, i.e. $t\gtrsim \mu/(B\rho^3g^3)^{1/4}$. Note also that relaxing the assumption of fastest growth at each instant could lead to a logarithmic correction as in Ref.~\cite{Kodio2017}.

The coarsening dynamics given by Eq.~(\ref{lambda-coarsening}) is different from the dynamics observed when the liquid layer is thin enough and where $\lambda\sim H^{1/2} t^{1/6}$~\cite{vandeparre2010hierarchical,Kodio2017}. To bridge the gap between both results and to determine the conditions under which coarsening dynamics applies, we have extended the derivation presented above to an arbitrary value of $H$, see Supplemental Material~\cite{SM}. For a shallow liquid bath such that $H/\lambda_S \lesssim 1/2$, the initial wavelength length scales as $\lambda_0/\lambda_S \sim (H/\lambda_S)^{1/2} (V/V_c)^{-1/6}$ while the wavelength grows as $\lambda/\lambda_S \sim (H/\lambda_S)^{1/2}\, \bar{t}^{\, 1/6}$, where $\bar{t}= B k_S^3 t/2\mu$.

\textit{Conclusions} -- In this Letter, we have shown from experiments and theoretical modeling that increasing the speed at which a sheet resting on a liquid is compressed/indented can drastically decrease the wavelength of the emerging wrinkles in a universal way independent of the geometry of the sheet. This procedure allows  patterns to be created at scales much smaller than those reachable through quasistatic confinement. In addition, we have shown that, when the confinement stops after the emergence of the wrinkles, their wavelength coarsens following a well defined temporal power law controlled by the sheet properties and the viscosity of the liquid. The exponent of this power law can even be changed according to the thickness of the liquid bath.

Although developed and validated using dedicated laboratory experiments, such as uniaxial compression and indentation of thin sheets on viscous fluids, the proposed model captures a generic viscous–elastic coupling. It is therefore expected to apply broadly to systems in which elastic thin films undergo dynamic compression while interacting with viscous environments. As an application, we briefly show that the proposed model can give some insight into a geological setting in which wrinkle-like features have been observed: the formation of rope-like structures (\emph{p\={a}hoehoe}) from lava flow~\cite{Fink1978,Griffiths2000,Slim2009}. Wrinkled \emph{P\={a}hoehoe} typically forms when flow occurs in a channel as the crust solidifies. While this uniaxial flow and compression is similar to the 1D case studied here, the solidification of the crust means it is difficult to make a precise comparison with the theory presented here and the wavelength that is observed. However, it is reasonable to ask whether the wavelength of the rope-like structure is likely to be close to the equilibrium wavelength $\lambda_S$, or influenced by the dynamics. For this purpose, we consider that the flow velocity is solely due to gravity: $V = H^2\rho g \sin \theta/(2\mu)$~\cite{Jeffreys1925,PGG2004,Farrell2018} (taking into account the outpouring velocity from the erupting vent would yield an even larger speed and increase the dynamical effects). Substituting this expression of $V$ in Eq.~(\ref{lambda0-asymp}), we get an universal relation for the wavelength,
\begin{equation}
\label{lam0-lava}
\frac{\lambda_0}{\lambda_S} = \left[\frac{4\pi}{3 \sin \theta}\right]^{1/3} \left[\frac{H^2}{h^2} \frac{\lambda_S}{L}\right]^{-1/3},
\end{equation}
where we used the explicit expression (\ref{lambda0}) of $V_c$. Using the explicit expression of $\lambda_S$, Eq.~(\ref{lam0-lava}) allows to compute the crust thickness when all other parameters are known. Using typical estimates $E= 100\mathrm{~GPa}$, $\nu = 0.3$, $\rho= 3000$ kg/m$^3$, $L=10$ m, $H=1$ m, $\theta = 5^\circ$ and $\lambda_0 = 0.1$ m \cite{Fink1978,Griffiths2000,Schilling2003}, we have $h \simeq 1.4$ mm. The quasistatic wavelength, $\lambda_S$, corresponding to this sheet thickness is close to 1 m, i.e. much larger than the observed value. This shows that the wavelength is strongly influenced by the dynamical effects. Indeed, $V/V_c \simeq 2500$ in this case and Fig.~\ref{fig03}(a) shows that $\lambda_0/\lambda_S \simeq 0.1$.

\begin{acknowledgments}
The authors acknowledge fruitful discussions with B. Davidovitch and E. Rapha\"el and the support by F.R.S.-FNRS under the research Grant (PDR ``ElastoCap'') No. T.0025.19. 
\end{acknowledgments}

\bibliographystyle{apsrev4-2}
\bibliography{dyn-wrinkle-refs.bib}

\end{document}